\documentclass[a4paper,12pt]{article}
\usepackage{shadow,pifont,epsfig,wrapfig,times,graphics,color} 
\usepackage{latexsym}

\begin{document}

\title{The Hypercentral Constituent Quark Model}
\author{M. M. Giannini, E. Santopinto, A. Vassallo\\
{\small Dipartimento di Fisica dell'Universita' di Genova and I.N.F.N. 
Sezione di Genova}}
\maketitle
\abstract{
The hypercentral constituent quark model contains a spin independent 
three-quark interaction inspired by lattice QCD calculations which reproduces the average energy of $SU(6)$ 
multiplets \cite{pl}. The splittings are obtained with a residual generalized $SU(6)$-breaking interaction 
including an isospin dependent term \cite{vass}. The long standing problem of the Roper is absent and all the
 3- and 4-star states are well reproduced.
The model has been used in a systematic way for transverse and longitudinal electromagnetic transition form 
factor of the 3- and 4- star and also for the missing resonances.
The prediction of the electromagnetic helicity amplitudes agrees quite well  with the
data except for low $Q^2$, showing that it can supply a realistic set of quark
wave functions. In particular we report the calculated helicity amplitude $A_{1/2}$
for the $S11(1535)$, which is in agreement with the TJNAF data \cite{dytman}.}

\section{Introduction}
The baryon  spectrum is usually described quite well by different Constituent 
Quark Models \cite{pl,is,sc}, which have in common the three 
constituent quark spatial 
degrees of freedom and the underlying SU(6) spin-flavour structure.
However they differ for the chosen potential, that is for the wave functions, and in particular also for 
the way in which the $SU(6)$ symmetry is broken.
By comparing various models one can see that each of them has different predictions concerning the position and number of the missing resonances under $2~GeV$, giving the possibility of discriminate among them, but 
due to the difficulties of the analysis of the experimental data this is not always possible.
In this respect the study of hadron spectroscopy is not sufficient to distinguish among the various forms of quark dynamics and to
this end one has to consider other observables more sensitive to the 
wave functions and to the way in which  SU(6) is broken such as the e.m. 
transition form factors, the so called helicity amplitudes.
\section{The Model}
The hypercentral Constituent Quark Model (hCQM) consists of a hypercentral quark interaction containing a
linear plus coulomb-like term, as suggested
by lattice QCD calculations \cite{bali}
\begin{equation}\label{eq:pot}
V(x)= -\frac{\tau}{x}~+~\alpha x~~,
 ~~\mathrm{with} ~~
~x=\sqrt{\mbox{\boldmath{$\rho$}}^2+{\mbox{\boldmath{$\lambda$}}}^2} ~~, 
\end{equation}
\noindent
where $x$ is the hyperradius defined in terms of the standard Jacobi
coordinates $\mbox{\boldmath{$\rho$}}$ and $\mbox{\boldmath{$\lambda$}}$.
We can think of this potential both as a two body potential in the 
hypercentral 
approximation (which, as shown by Fabre de la Ripelle \cite{fdlrip},is a good approximation for the lower 
states) or as a true three body potential.
 A hyperfine term of the standard
form \cite{is} is added and treated as a perturbation.
The parameters $\alpha$, $\tau$ and the strength of the
hyperfine interaction are fitted to the spectrum ($\alpha=1.61~fm^{-2}$, 
$\tau=4.59$ and the strength of the
hyperfine
interaction is determined by the $\Delta$ - Nucleon mass difference).
Having fixed the values of these parameters, the resulting wave functions   
have been used for the calculation of the
photocouplings \cite{aie}, the transition form factors for the negative parity resonances 
 \cite{aie2}, the elastic form factors
\cite{mds} and the ratio between the electric and magnetic form factors of
the proton \cite{rap} and now also for the longitudinal and transition form factors for 
all the 3- and 4-stars and the missing resonances \cite{tobepu}.
\section{The electromagnetic transition form factors}
The helicity amplitudes for the electroexcitation of baryon resonances,
$A_{1/2}$, $A_{3/2}$ and $S_{1/2}$ are calculated as the transition matrix elements
of the transverse and longitudinal electromagnetic interaction between the
nucleon and the resonance states given by this model. A non relativistic 
current for point quarks is used.
\begin{figure}
\begin{center}
\includegraphics[height=8cm,width=5cm,angle=90]{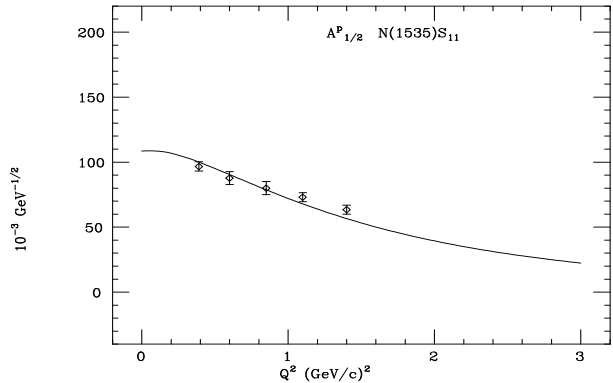}
\caption{Comparison between the new TJNAF data for the helicity 
amplitudes $A^p_{1/2}$ for the  $S_{11}(1535)$ \cite{dytman} 
and the calculations with the hCQM\cite{aie}.}
\end{center}
\end{figure}   
\noindent 
In particular, in Fig.1, one can see the predictions (full line) for 
the $A^{p}_{1/2}$ for the $S11(1535)$ \cite{aie2} 
which are in agreement with the new TJNAF data \cite{dytman}. 
The results for the helicity amplitudes for all the negative parity
resonances are reported in Ref.\cite{aie2}. The prediction of the electromagnetic helicity amplitudes agrees quite well  with the
data  showing that the hCQM can supply a realistic set of quark
wave functions.
In general the $Q^2$ behaviour is reproduced, except for
discrepancies at small $Q^2$, specially in the
$A^{p}_{3/2}$ amplitude of the transition to the $D_{13}(1520)$ state.
These discrepancies could be ascribed to the non-relativistic character of
the model, and to the lack of explicit quark-antiquark configurations
which may be important at low $Q^{2}$.
The longitudinal and transverse transition form factors for 
all the 3- and 4-stars and the missing resonances have been calculated \cite{tobepu}.
 The computer code is at 
disposal under request and it can be used also with other models. 
In this way it can be a useful tool for 
the forthcoming analysis of the experimental data.

The relativistic corrections at the level of boosting the nucleon and the
resonances states to the EVF or the Breit frame are important but not sufficient \cite{mds2}.
These boost effects are indeed important for the elastic e.m. form
factors, giving origin to the decreasing behaviour of the ratio R of the electric
and magnetic proton form factors, as shown by the recent TJNAF data
\cite{ped}.
\section{$SU(6)$-breaking residual interaction.}
There are different motivations for the
introduction of a residual flavour dependent term in the three-quark 
interaction.
The well known Guersey-Radicati mass formula \cite{gura}
contains a flavour dependent term, which is essential for the description  
of the strange spectrum.
In the chiral Constituent Quark Model \cite{ple}, the non
confining part of the
potential is provided by the interaction with the Goldstone bosons,
giving rise to a spin- and isospin-dependent interaction.
More generally, one can expect that the quark-antiquark pair production
can lead to an effective residual quark interaction containing an isospin
(or
flavour) dependent term and with these motivations in mind, we have 
introduced \cite{vass} isospin 
dependent terms in the hCQM hamiltonian.
The complete interaction used is given by
\begin{equation}\label{tot}
H_{int}~=~V(x) +H_{\mathrm{S}} +
H_{\mathrm{I}} +H_{\mathrm{SI}}~,
\end{equation}
were V(x) is the linear plus hypercoulomb SU(6)-invariant potential,
already described in Eq.(1),
 while $H_{\mathrm{S}} +
H_{\mathrm{I}} +H_{\mathrm{SI}}$, is a residual SU(6)-breaking
interaction that can be treated as a perturbative term leading to an  
improved description of the spectrum.   

\section{Conclusions}
We have presented various results predicted by 
 the hypercentral Constituent Model compared with the
experimental data.
We have shown that the hCQM can supply a realistic set of quark wave 
functions.

\end{document}